# Geometric Filterless Photodetectors for Mid-infrared Spin Light

Jingxuan Wei[1], Yang Chen[2], Ying Li[3,4], Wei Li[5], Junsheng Xie[1,6], Chengkuo Lee[1,6]*, Kostya S Novoselov[7], Cheng-Wei Qiu[1]*

[1]Department of Electrical and Computer Engineering, National University of Singapore, Singapore 117583, Singapore.

[2]CAS Key Laboratory of Mechanical Behaviour and Design of Materials, Department of Precision Machinery and Precision Instrumentation, University of Science and Technology of China, Hefei 230026, China.

[3]Interdisciplinary Center for Quantum Information, State Key Laboratory of Modern Optical Instrumentation, ZJU-Hangzhou Global Scientific and Technological Innovation Center, Zhejiang University, Hangzhou 310027, China.

[4]International Joint Innovation Center, Key Lab of Advanced Micro/Nano Electronic Devices & Smart Systems of Zhejiang, Zhejiang University, Haining 314400, China.

[5]GPL Photonics Lab, State Key Laboratory of Applied Optics, Changchun Institute of Optics, Fine Mechanics and Physics, Chinese Academy of Sciences, Changchun 130033, China.

[6]Center for Intelligent Sensors and MEMS, National University of Singapore, Singapore 117608, Singapore.

[7]Institute for Functional Intelligent Materials, National University of Singapore, Singapore 117544, Singapore.

* E-mail: (C. L.) elelc@nus.edu.sg and (C.-W. Q.) chengwei.qiu@nus.edu.sg

# Abstract

Free-space circularly polarized light (CPL) detection, requiring polarizers and waveplates, has been well established, while such spatial degree of freedom is unfortunately absent in integrated on-chip optoelectronics. So far, those reported filterless CPL photodetectors suffer from the intrinsic small discrimination ratio, vulnerability to the non-CPL field components, and low responsivity. Here, we report a distinct paradigm of geometric photodetectors in mid-infrared exhibiting colossal discrimination ratio, close-to-perfect CPL-specific response, a zero-bias responsivity of 392 V/W at room temperature, and a detectivity of ellipticity down to $0.03°$ $Hz^{-1/2}$. Our approach employs plasmonic nanostructures array with judiciously designed symmetry, assisted by graphene ribbons to electrically read their near-field optical information. This geometry-empowered recipe for infrared photodetectors provides a robust, direct, strict, and high-quality solution to on-chip filterless CPL detection and unlocks new opportunities for integrated functional optoelectronic devices.

# Introduction

Polarization, as another dimension of light beyond the intensity and wavelength, plays a pivotal and ubiquitous role in various optical technologies[1–3]. In particular, circularly polarized light (CPL), also known as spin light that carries spin angular momentum, enables widespread applications such as drug screening[4], quantum optics[5], imaging[6], and biosensing[7,8]. Conventionally, the detection of CPL requires a set of bulky polarization optics because the commercial photodetectors are only sensitive to the light intensity, imposing difficulties for miniaturized systems. One possible solution is flat optics, where the filters and wave plates can be replaced with ultrathin metasurfaces correspondingly[9–12]. However, this recipe still indispensably relies on a propagation distance in space between metasurfaces and detectors[1,13,14], multiple measurements to disentangle different polarization states, an accurate alignment and heterogeneous integration[15,16], preventing itself from the implementation of cost-effective monolithic systems. That leads to the development of novel photodetectors natively sensitive to the spin state of light[17–19]. So far, to the best of our knowledge, there are three main approaches to designing spin-sensitive photodetectors. First, chiral molecules with circular dichroism can be used in the phototransistor configuration for high-responsivity CPL detection[17,19–21]. However, the discrimination ratio of the photoresponse to left-handed and right-handed CPL (i.e., LCP and RCP), $g_{ph} = |(I_{LCP} − I_{RCP})/(I_{LCP} + I_{RCP})|$, is intrinsically small because of the weak chiroptic response in molecules. Besides, these devices are usually vulnerable to unpolarized and linearly polarized light. Second, artificial plasmonic structures can provide larger circular dichroism, but an efficient way of transferring the chiroptical response to an electrical signal readout is still lacking[18,22–24]. Third, CPL-sensitive photoresponse also exists in circular photogalvanic effect[25–27], photon drag effect[28,29], and spin-galvanic effect[30], but those responsivities are still too low for most practical applications.

Practically speaking, an effective monolithic CPL detector ought to meet the following three criteria: (i) the left-handed and right-handed CPL can be well discriminated[17,27]; (ii) the detection of CPL is robust, with immunity against the ubiquitous unpolarized and linearly polarized light[24,31]; (iii) the responsivity and signal-to-noise ratio should be sufficiently large for practical applications[6]. The criteria mentioned above can also be mathematically formulated. We first write the general expression for a polarization-dependent photoresponse: $V_{ph} = R_0 \cdot S_0 + R_1 \cdot S_1 + R_2 \cdot S_2 + R_3 \cdot S_3$, with the $R_{0,1,2,3}$ as the respective responsivities to the four Stoke parameters, $S_{0,1,2,3}$ (see supplementary note S1 for detailed discussion). Note that $S_0$, $S_1$, $S_2$, and $S_3$ represent the intensity of light, two linearly polarized components, and the circularly polarized light component, respectively[6]. A perfect CPL photodetector should follow:

$$V_{ph} = R_3 \cdot S_3 \tag{1}$$

meaning that $R_3$ dominates in the photoresponse over $R_{0,1,2}$. Notably, the above equation also indicates that LCP ($S_3 = -1$) and RCP ($S_3 = 1$) light will lead to opposite photoresponse so that they can be markedly discriminated by the photovoltage sign regardless of their intensities. The concept of such an ideal CPL-specific detector can be better visualized in the Poincaré sphere, as shown in **Fig. 1a**. However, it remains elusive to realize such rigid CPL photodetectors with high performance.

In this study, we showcase how the symmetry and geometry of plasmonic nanostructures array, as well as patterned graphene ribbons, come into a synergetic play to enable an on-chip specific detection of CPL with a colossal discrimination ratio, immunity to non-CPL field components, and high responsivity. Our contribution is three-fold: first, we reveal that the vectorial photoresponse generated from achiral structures could reach a symmetry-protected infinite discrimination ratio, $g_{ph}$, thereby superior to the existing works based on circular dichroism whose $g_{ph}$ is naturally constrained below one[17–19]. Second, we demonstrate how the geometric arrangement of plasmonic nanostructures can power up the specific detection of CPL. Third, we report how the patterned graphene sheet enhances the efficiency in reading the near-field information of the plasmonic nanostructures. Our graphene ribbons device reaches a peak responsivity of 392 V/W for the individual nanostructures and a detectivity of ellipticity down to 0.03° $Hz^{-1/2}$ in the mid-infrared at room temperature.

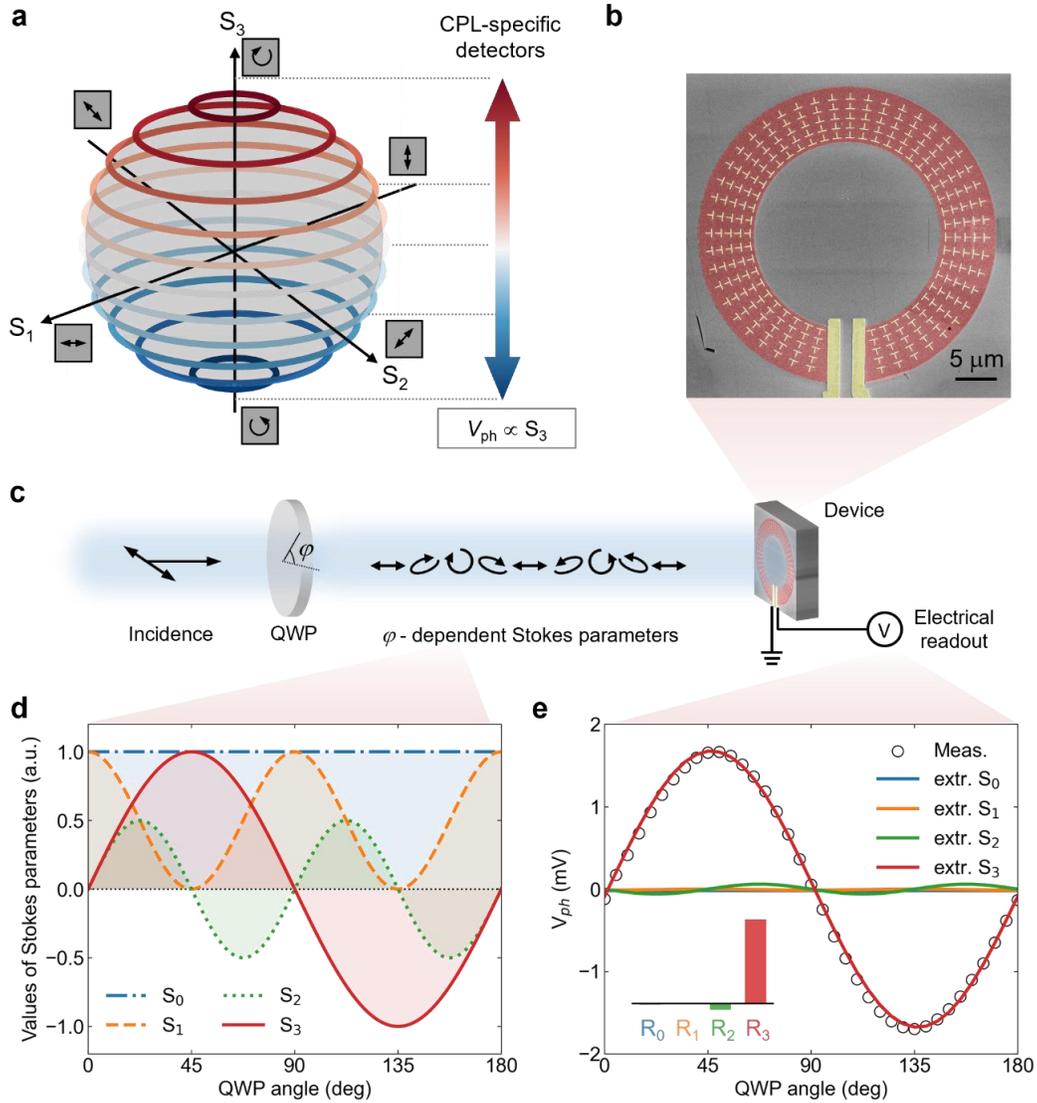

**Fig. 1 | Concept of geometric photodetectors for CPL-specific detection. a**, Illustration of CPL-specific photodetectors in the Poincaré sphere, where the photoresponse, $V_{ph}$, only depends on the fourth Stokes parameter of the incident light, $S_3$. **b**, False-coloured scanning electron microscopy image of our ring-shaped device, consisting of geometrically arranged mirror-symmetric plasmonic nanostructures on a graphene sheet. Colour code: graphene in red; plasmonic nanostructures and electrodes in yellow; substrate in grey. **c**, Schematic experimental setup and methodology. The responsivities of our device to the four Stokes parameters are studied by rotating a quarter-wave plate (QWP) and then reading the electrical output of our device. **d**, Calculated dependence of Stokes parameters on the QWP angle. **e**, Experimentally measured QWP angle-dependent photovoltages and the extracted contributions from $S_{0,1,2,3}$, under a normally incident illumination at 4 μm wavelength. Inset shows the extracted responsivities, $R_{0,1,2,3}$, confirming the dominance of CPL sensitivity in our device. The incidence power is 0.9 mW. Note that the handedness of CPL in this work is defined from the point of view of the source.

# Results

## Methodology

**Figure 1b** illustrates the false-colored scanning electron microscopy (SEM) image of one design of our CPL-specific photodetectors. This two-port device features a ring-shaped few-layer graphene device channel and circularly arranged metallic T-shaped nanoantennas array (5/60 nm Pd/Au). See Fig. S4 for the atomic force microscopy measurement. A gap is created in the ring-shaped channel to place the electrodes. The device is fabricated on a silicon wafer with 285 nm thick thermal oxide. Devices with other geometries such as half ring and L-shape will be discussed later.

The methodology of our experiment is depicted in **Fig. 1c**. The responsivities of our photodetectors to the four Stoke parameters were studied by modulating the incident beam with a quarter-wave plate (QWP). Given that the incident light is purely coherent and linearly polarized along $x$-axis, the polarization state of light after passing the QWP can be described with a new set of Stokes parameters (see supplementary note S2 for derivation): $S_0 = 1$; $S_1 = 0.5+0.5\cos(4\varphi)$; $S_2 = 0.5\sin(4\varphi)$; $S_3 = \sin(2\varphi)$, where $\varphi$ is the angle of the fast axis of the QWP (**Fig. 1d**). Because the four Stokes parameters show non-synchronous dependence on the QWP angle, their respective contribution to the photoresponse in our device can be extracted. The incident light is at 4 μm wavelength and normal to the device. In our experiments, we always use the open-circuit voltages as the photoresponse unless otherwise stated. **Figure 1e** shows the measured $\varphi$-dependent photoresponse, following the $\sin(2\varphi)$ function nicely. From the general expression of QWP angle-dependent photovoltage: $V_{ph} \propto R_0 + R_1 \cdot (0.5+0.5\cos(4\varphi)) + R_2 \cdot 0.5\sin(4\varphi) + R_3 \cdot \sin(2\varphi)$, we can extract the respective normalized responsivities: $R_{0,1,2,3} \sim (-0.013, -0.006, -0.072, 1)$, confirming that $R_3$ is ten times larger than the sum of $R_0$, $R_1$, and $R_2$. Notably, the discrimination ratio, $g_{ph}=|R_3/R_0|=84$, even breaks the limit of one in conventional CPL detectors based on circular dichroism[17–19]. Therefore our device can be concluded to meet the first two criteria of an ideal CPL-specific photodetector. The measured $R_3$ at room temperature reaches 1.8

V/W (see Fig. S5 for the estimation of incident power). Much larger responsivities can also be achieved, as discussed later.

## Mirror-symmetric meta-atoms with colossal discrimination ratio

The design principle of our proposed plasmonic nanostructures is illustrated in **Fig. 2a**. Since CPL possesses handedness, it is commonly deemed that CPL-sensitive photodetectors need to break the mirror symmetry in geometry[32–34]. However, this perception will be challenged by closer scrutiny on the symmetry analysis. Although the mirror symmetry in achiral structures prohibits a different absorption between LCP and RCP illumination, there could exist CPL-sensitive vectorial photoresponse, e.g., $J_x$, which is parity-odd to the reflection operation. More interestingly, the achiral symmetry forces the $J_x$ under LCP and RCP illumination to be equal in magnitude but with opposite signs, so that the photoresponse discrimination ratio, $g_{ph} = |(J_{x,LCP} - J_{x,RCP})/(J_{x,LCP} + J_{x,RCP})|$, will approach infinity, in sharp contrast to the absorption-based chiral devices whose $g_{ph}$ is constrained below one[17–19]. Further analysis reveals that the symmetry-protected infinite $g_{ph}$ should be only available when the nanostructures contain a single mirror symmetry axis (see supplementary note S3).

To implement the vectorial photoresponse with colossal $g_{ph}$, we turn to study the near-field of nanophotonic structures. Recent works have revealed a giant chirality "hidden" in the near-field profile of mirror-symmetric achiral nanostructures via cathodoluminescence microscopy, which originates from the interference of multiple modes[35–38]. However, the efficient reading of the near-field chirality with an electrical output remains elusive and unexplored. In our preliminary design, we realize the transferring process by placing an unpatterned graphene sheet below the plasmonic nanostructures, so that the asymmetric near-field can generate directional nonlocal photoresponse at the metal-graphene interfaces[39–41]. The dominant microscopic photoresponse mechanism is the widely accepted hot carriers-assisted photo-thermoelectric effect in graphene[42–44], with its unique directional nonlocal feature supported by our nanometer-resolution photovoltage mapping experiment (Fig. S6). We also note that intrinsic graphene plasmons should not play an essential role in our device because they cannot be efficiently excited at this wavelength (Fig. S7). Detailed theoretical analysis on the near field asymmetry and the resulting CPL-sensitive photoresponse is provided in supplementary note S4.

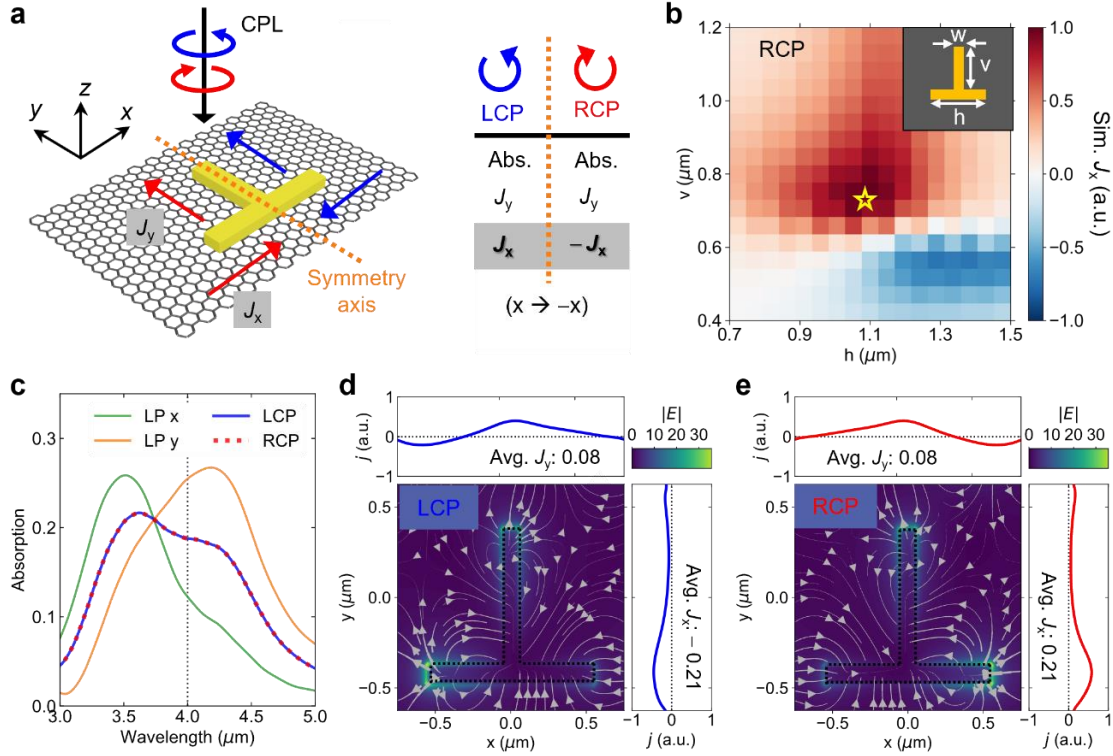

**Fig. 2 | Mirror-symmetric meta-atoms with colossal discrimination ratio. a**, Symmetry analysis of the photoresponse in our designed achiral plasmonic nanostructure which sits on a graphene sheet. $J_x$ and $J_y$ denote the vectorial photocurrents along the $x$ and $y$ direction, respectively. LCP: left-handed CPL; RCP: right-handed CPL. Under the reflection operation ($x \rightarrow -x$), the absorption of nanostructure and $J_y$ are parity-even, and hence respond identically to LCP and RCP. In contrast, $J_x$ is parity-odd and thus CPL-sensitive. **b**, Calculated $J_x$ under RCP illumination in the parameter space of the nanostructure consisting of two vertical and horizontal bars. For a fixed bar width, $w$, of 0.1 μm, the optimal CPL response is achieved at $h = 1.1$ μm and $v = 0.75$ μm, denoted by the hollow star. **c**, Simulated absorption spectra of the optimized plasmonic nanostructures under linearly and circularly polarized illumination. As expected, the achiral nanostructures show no circular dichroism in the far field. **d,e**, Simulated near-field profile and corresponding photocurrent flow in graphene of the optimized plasmonic nanostructures under the LCP (**d**) and RCP (**e**) illumination, respectively. Grey lines denote the current flow. When the handedness of CPL changes, the averaged $J_y$ remains while the averaged $J_x$ flips its sign. The resultant photoresponse discrimination ratio, $g_{ph}$, will then be infinite, protected by its achiral geometry.

We further optimize the geometric parameters of the plasmonic nanostructures by numerically calculating the CPL-sensitive $J_x$ (see Fig. S8 for the simulation details). **Figure 2b** shows the results at different horizontal lengths, $h$, and vertical lengths, $v$, of the nanostructures under RCP illumination. The $J_x$ under LCP illumination differs by a negative sign. The width, $w$, and the thickness of the nanostructures are kept as 0.1 μm and 0.06 μm, respectively. The $x$-axis and $y$-axis pitches of the nanostructures array in our design are 1.6 μm. The optimized dimensions of the nanostructure are $h = 1.1$ μm and $v = 0.75$ μm throughout this work, as denoted with the hollow star in the figure. The calculated result has been validated experimentally, where we fabricated and characterized a group of control devices with different geometries (Fig. S9).

To better illustrate our design principle, we plot the simulated absorption spectra, near-field profile, and the resulting CPL-sensitive photoresponse of the optimized geometry obtained in Fig 2b. The achiral nanoantenna identically absorbs the LCP and RCP light, thereby showing no circular dichroism (**Figure 2c**). Notably, there is a wavelength shift in the absorption peaks under the $x$-axis and $y$-axis linearly polarized illumination, which usually is a feature of the chirality-dependent mode interference and hence the near-field chirality (see supplementary note S4). The near-field analysis for LCP and RCP illumination are illustrated in **Fig. 2d and Fig. 2e**, respectively. The near-field distribution is highly asymmetric, with an intensity ratio of 6 (Fig. S10). When the handedness of CPL changes, the field profile flips horizontally, leading to opposite average $J_x$ and identical average $J_y$ (see Fig. S11 for experimental validation). Therefore, we have theoretically realized the symmetry-protected infinite $g_{ph}$ in achiral structures with no circular dichroism, fulfilling the first criterion of ideal CPL detectors.

## Geometrically arranged meta-atoms with immunity to non-CPL field components

Moving forward, the second criterion requires our device to be specific to CPL, namely, immune against both unpolarized and linearly polarized light. By analyzing the simulated polarization dependence of the $J_x$ (Fig. S12), we can extract the respective normalized responsivities to the four Stokes parameters as $(R_0, R_1, R_2, R_3) \sim (0, 0, -0.18, 0.21)$. The vanished $R_0$ means that $J_x$ has been immune to the polarization-insensitive term, as a natural result of the mirror-symmetric geometry of our design. However, the nonvanishing $R_2$ indicates that our design is still affected by linearly polarized light. Such dependence on linearly polarized light ubiquitously exists in all materials and structures except the isotropic substances with optical activity[45], which, unfortunately, in turn, forbid the vectorial photoresponse in a planar device (supplementary note S3).

To address the above bottleneck, we propose the geometric arrangement of nanostructures to realize a buildup in CPL response and cancellation in the linearly polarized light, which is inspired by the geometric metasurfaces where the nanostructures are rotated in a non-uniform manner to accumulate different geometric phases for the selective steering of polarization states[14,38,46]. The core of our method is the different polarization dependence of oriented nanostructures, with the experimental schematic shown in **Fig. 3a**. In our experiments, we rotated a rectangular device by an angle of $\theta$ with a step of 15º and measured the photovoltages (Fig. S13, and **Figure 3b** shows the results with a step of 30º for clarity). The extracted $\theta$-dependent Stokes responsivities, $R_{0,1,2,3}$, are shown in **Fig. 3c**. As we have discussed, the $R_0$ simply vanishes everywhere. Notably, the $R_1$ ($\sim \sin(2\theta)$) and $R_2$ ($\sim -\cos(2\theta)$) vary cyclically, resembling a geometric phase-like behavior. In contrast, $R_3$ is rotation-invariant. The above results can also be understood by the rotational symmetry of the defined Stokes parameters. Therefore, we can achieve specific CPL detection using geometrically arranged nanostructures. Such a conclusion is also supported by our measurements on the segments of a curved device (Fig. S14).

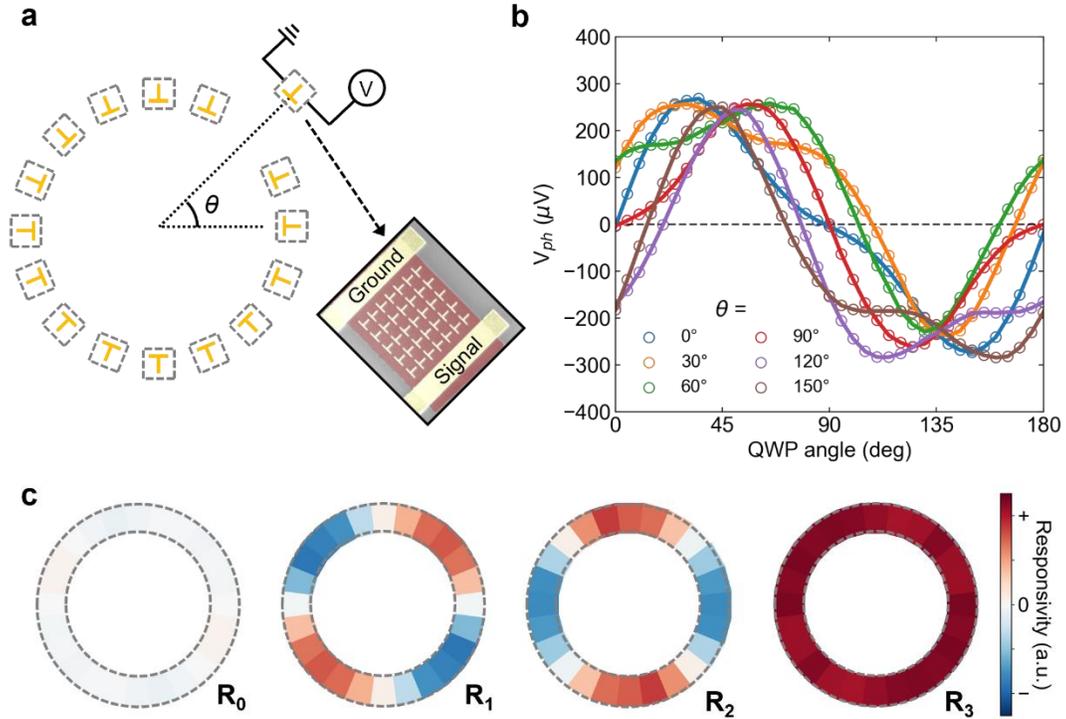

**Fig. 3 | Rotational symmetry of the polarization-dependent responsivities in meta-atoms. a**, Schematic of our approach to characterize the orientation-controlled polarization dependence of the plasmonic nanostructure. The nanostructure is rotated along with the electrodes at an angle of $\theta$. Inset shows the used rectangular device in our experiments. **b**, Measured polarization-dependent photovoltages at different $\theta$. **c**, Extracted $R_{0,1,2,3}$ at different $\theta$. Notably, $R_0$ vanishes everywhere due to the mirror symmetry of meta-atoms. $R_1$ and $R_2$ follow the functions of $\sin(2\theta)$ and $-\cos(2\theta)$, respectively. $R_3$ is rotation invariant. Therefore, the different rotational symmetries of $R_{0,1,2,3}$ of the meta-atoms allow us to achieve CPL-specific detection via geometrical arrangement.

In this work, we have demonstrated three combination types, namely, (i) ring-shaped cascaded device with $\theta \in (-80^o, 260^o)$, where $R_1 \propto \int_{\theta=-80^o}^{260^o} \sin(2\theta) d\theta \approx 0$ and $R_2 \propto -\int_{\theta=-80^o}^{260^o} \cos(2\theta) d\theta \approx 0$, as already shown in Fig. 1e; (ii) half ring-shaped cascaded device with $\theta \in (0^o, 180^o)$, where $R_1 \propto \int_{\theta=0^o}^{180^o} \sin(2\theta) d\theta = 0$ and $R_2 \propto -\int_{\theta=0^o}^{180^o} \cos(2\theta) d\theta = 0$, shown in Fig. S15; (iii) L-shaped cascaded device with $\theta \in 0° \& 90°$, where $R_1 \propto \sum_{\theta=0°,90°} \sin(2\theta) = 0$ and $R_2 \propto -\sum_{\theta=0°,90°} \cos(2\theta) = 0$, shown later. In principle, the open-circuit voltage and the short-circuit current only scale with the row number and column number of nanostructures cascaded in series and parallel, respectively, regardless of the particular combination type. In the viewpoint of practical application, the ring- and half-ring-shaped devices suffer larger footprints and more complicated structures arrangement, thereby inferior to the L-shaped cascaded device.

## Enhanced CPL-specific detection using graphene ribbons

We turn to the third criterion: high-responsivity CPL detector. Graphene has long been mentioned for its low photoresponse efficiency due to its semi-metallic nature[47]. However, little attention has been paid to the role of geometry in graphene photodetectors. Intuitively, the ubiquitous graphene sheet could form a short circuit around the nanostructures, reducing the external output. Closer scrutiny reveals that our design only requires a small graphene area to convert the optical near-field to voltages and convey the electrical signals to the external circuit. Therefore, we may play with the geometry of the graphene layer to increase the photoresponse. As shown in **Fig. 4a**, we fabricated and compared two L-shaped cascaded devices, which use graphene sheets and ribbons to extract the near-field chirality of nanostructures, respectively. To guarantee a fair comparison, we fabricated the two devices out of the same graphene flake and in the same fabrication process. The ribbons width in our experiments was chosen to be 0.6 μm, as a compromise to our fabrication and alignment error. Our simulation shows that the CPL-sensitive photoresponse will not be affected, as long as the graphene ribbons cover the left and right ends of the T-shaped nanostructure (**Fig. 4b**). Besides, the reduced channel width can better guide photoresponse flow, suppressing the excessive loss in the graphene sheet. Our experimental results show that the photoresponse of the ribbons device is five times large as the sheet device, while the two devices have the same device area (**Fig. 4c**). This enhancement was repeated in another two pairs of devices (Fig. S16). At an incident power of 45 μW and zero external bias, the extracted responsivities of the ribbons device are $R_{0,1,2,3} \sim (0.03, 0.08, 0.08, 1) \times 15$ V/W, corresponding to a $g_{ph}$ of 33. Considering that our device consists of four parallel-connected columns which increase the photocurrent response but not photovoltage response, the photovoltage responsivity of individual nanostructures is 60 V/W.

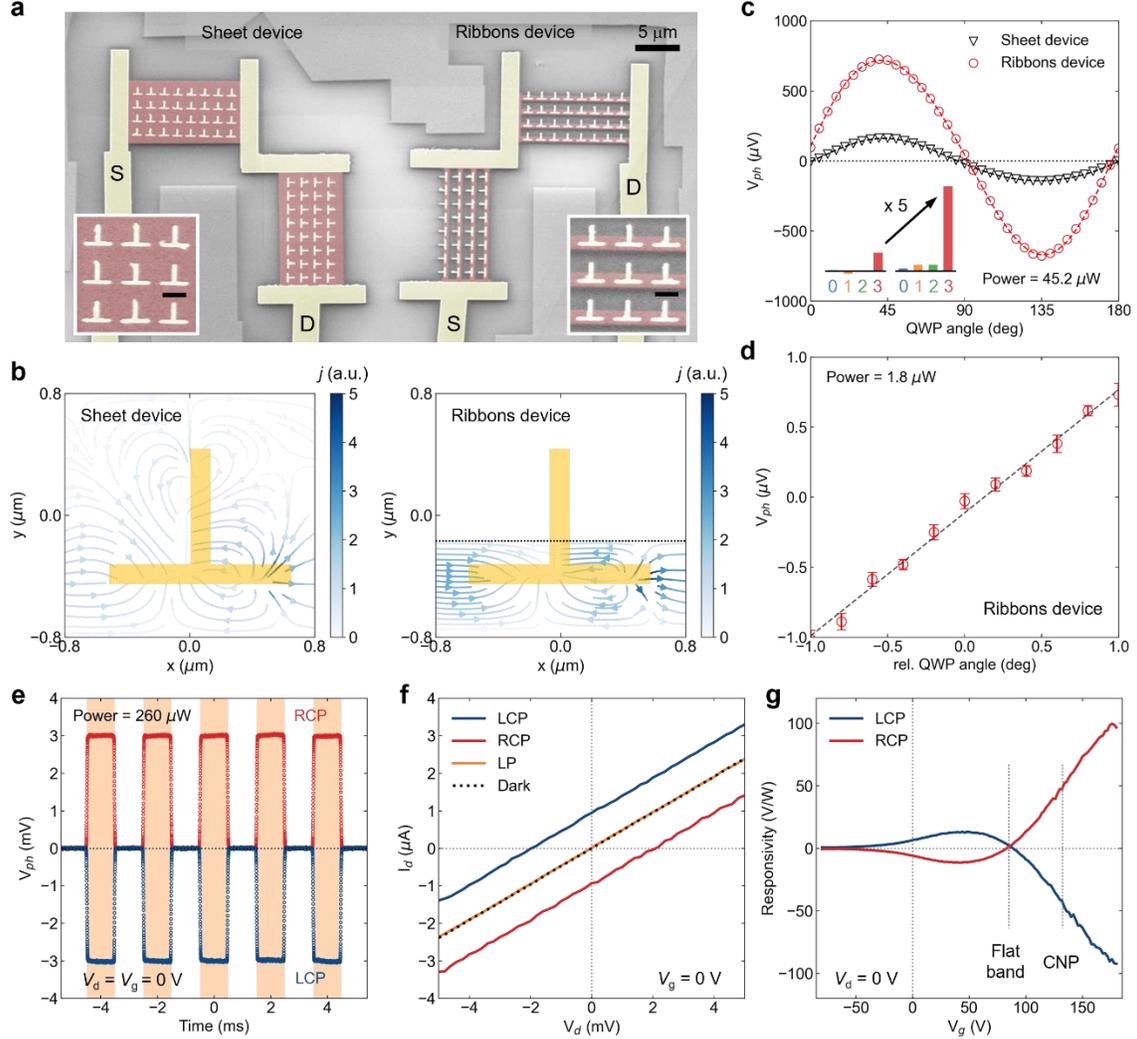

**Fig. 4 | Enhanced CPL-specific detection using graphene ribbons. a**, SEM images of the fabricated graphene sheet and ribbons devices. Scale bars in insets are 1 μm. **b**, Simulated current flow in the graphene sheet and ribbons devices. After removing the excessive graphene, the ribbons device can guide the photocarriers better and possesses a larger internal resistance, thereby reducing the internal energy loss and increasing the external electrical output. **c**, Measured polarization-dependent photovoltages in sheet and ribbons devices. Insets show the extracted responsivities, $R_{0,1,2,3}$, of the sheet device (left) and the ribbons device (right). Both devices specifically respond to the CPL, and the use of ribbons enhances the responsivity by five times, reaching 15 V/W at zero external bias. **d**, Fine measurement of the optical ellipticity at a small incident power of 1.8 μW, showing a detectivity of 0.03° Hz$^{-1/2}$. **e**, Measured photovoltages during on-off cycles of LCP

and RCP light illumination. Orange shades indicate the light condition. **f**, Measured *I-V* curves with drain-source bias. The *x* and *y* axes intersections are open-circuit voltages and short-circuit currents. LP: Linearly polarized light. **g**, Measured gate-tunable photoresponse. No drain-source bias, $V_d$, was applied. CNP, charge neutral point of graphene. The peak responsivity of the four-column ribbons device reaches 98 V/W, corresponding to a responsivity of 392 V/W in single-column nanostructures.

Without applying any bias, our device then has no dark current and hence very low noise level. The dark noise density of the ribbons device is down to 10 nV Hz$^{-1/2}$ at working frequencies above 100 Hz, approaching its Johnson noise limit (Fig. S17), in agreement with previous zero-bias graphene devices[48]. The noise equivalent power (NEP) of our device in passive mode is then about 0.67 nW Hz$^{-1/2}$. In practice, the performance of CPL detectors is usually limited by the dominant light noise under illumination rather than the dark noise[49], which has been largely ignored in the literature. For a given incident power, *P*, the detectivity of optical ellipticity can be formulated as $D_{ellipticity} = (180°/\pi) \cdot N_{light} / (2 \cdot R_3 \cdot P)$, where $N_{light}$ is the power-dependent light noise. At a small incident power of 1.8 μW, our device can measure the ellipticity down to 0.03° Hz$^{-1/2}$ (**Fig. 4d**, see methods for experiment details). We note that a higher detectivity should be possible if the laser noise can be suppressed, because we have observed a saturated detectivity at higher incident power due to the dominance of laser noise (Fig. S18). Nevertheless, the demonstrated performance should already be sufficient for many applications such as particle identification[50], imaging and circular dichroism spectroscopy[4,51].

We further characterize the time response of the ribbon device. **Figure 4e** shows the measured photovoltages with an oscilloscope during on-off cycles of LCP and RCP light illumination, which was modulated by an optical chopper at 500 Hz. Zero external bias was applied ($V_d = V_g = 0$ V), and no preamplifier was used. We then used a pulsed laser with nanosecond duration to probe the rise and fall times of our device as 886 ns and 902 ns, respectively (Fig. S19). We note that the measured response speed seems not to be limited by the intrinsic speed of our device because we observed a linear relationship between the response time and the aspect ratio of our device (Fig. S20). Following the *RC* delay model, the measured response speed is most likely limited by an unknown parasitic capacitance of a few tens of pF in our setup. Our experiments also show almost constant photoresponse up to 4 kHz modulation frequency (Fig. S21). The measured photoresponse speed is already much faster than most mid-infrared detectors such as thermopiles and microbolometers[52–54].

We then measured the *I-V* characteristics of our device. As shown in **Fig. 4f**, the current is linearly dependent on the applied drain-source bias, indicating the Ohmic contact in graphene device. At CPL illumination, the *I-V* lines form two intersections on *x* and *y* axes, indicating the open-circuit voltage and short-circuit current, respectively. The *I-V* lines under LCP and RCP illumination are respectively shifted upwards and downwards, illustrating their opposite photoresponse. Thereby, linearly polarized light that equally combines LCP and RCP light does not show measurable photoresponse, as a feature of our CPL-specific detector. The photoresponse can be effectively tuned in both magnitude and sign by applying a gate voltage (**Fig. 4g**). The non-monochronic gate dependence is a feature of the photoresponse originating from graphene-metal interfaces[42,43]. Notably, the photoresponse vanishes at the gate voltage of 86 V, where the flat band condition is met at graphene-metal interfaces. The charge neutral point (CNP) is achieved at the gate voltage of 136 V (Fig. S22), indicating that Pd contact dopes the underneath graphene to be *p*-type. The hole mobility of our ribbon device was extracted to be 2446 $cm^2V^{-1}s^{-1}$. The peak responsivity of the four-column device reaches 98 V/W at the gate voltage of 180 V, corresponding to a responsivity of 392 V/W in the single-column nanostructures.

We also found that the photoresponse will be slightly saturated at high incident power due to the screening effect (Fig. S23), agreeing with the previous report[55]. The calculated wavelength dependence shows that the full width at half maximum (FWHM) of our current design is about 1 μm (Fig. S24).

## Discussion

This work conceptualizes and demonstrates geometry-empowered monolithic CPL-specific photodetectors with colossal discrimination ratio, immunity to non-CPL field components, and high responsivity, setting a benchmark for future integrated on-chip polarimeters. We reveal the importance of geometry in integrated photodetectors for both novel functionality and high responsivity. Although the bipolar CPL-specific detection has been reported in photogalvanic effect[31], ratchet effect[56], and thermopiles[24], their responsivities are at least four orders lower than our work, and a clear guideline is lacking. Besides, our demonstrated gate tunability is also a unique advantage, which is unavailable in conventional bulky materials. In the viewpoint of symmetry, the sign of photoresponse in our device can be flipped by the reversal of either parity (LCP to RCP), charge (holes to electrons), or time (momentum of charges). Since the near field of resonant nanostructures could display more information on the various attributes of light than the far-field scattering[57–59], our demonstrated efficient and specific readout of the near field with an electrical signal will be helpful for new functional optoelectronic devices. For example, our principle can be readily extended to the specific detection of other Stokes parameters, $S_{0,1,2}$ (see supplementary note S5). More exciting is that the geometric photodetectors are born to appropriately deal with the structured light such as vector beam and optical vortex whose intensity and phase profile are space-variant[60,61]. Higher responsivity is also possible by using a metallic substrate with a dielectric spacer to reduce the reflection loss[62] and narrower graphene ribbons. Our device is less than 100 nm thick and compatible with the complementary metal-oxide-semiconductor (CMOS) platform[63]. Thanks to the recent advances in wafer-scale single-crystal growth of graphene[64], our work provides new opportunities for integrated, cost-effective, functional optoelectronic devices.

## Acknowledgments

C.-W.Q. acknowledges financial support from the National Research Foundation (Grant No. NRF-CRP22-2019-0006) and Advanced Research and Technology Innovation Centre (Grant No. R-261-518-004-720). C.L. acknowledges financial support from the National Research Foundation Singapore (Grant No. NRF-CRP15-2015-02). Y.C. acknowledges the support from the start-up funding of the University of Science and Technology of China and the CAS Pioneer Hundred Talents Program. Y.L. acknowledges the support from the National Natural Science Foundation of China (Grant No. 92163123). W.L. acknowledges the financial support from the National Natural Science Foundation of China (Grant No. 62134009, 62121005) and the Innovation Grant of Changchun Institute of Optics, Fine Mechanics and Physics (CIOMP).

## Author contributions

J.W. and C.-W.Q. conceived the project. J.W. and Y.C. did the theoretical analysis and numerical simulation. J.W. fabricated the samples. J.W., J.X. and C.L. carried out and contributed to the device characterization. J.W., Y.C., Y.L., W.L., C.L., K.S.N., and C.-W.Q discussed and analyzed the numerical and experimental results. All authors discussed and contributed to the manuscript. C.-W.Q. oversaw the whole project.

## Competing interests

The authors declare no competing interests.

## Additional information

Supplementary information is available for this paper.

## Methods

**Simulation.** The near field profile and absorption of our designed nanostructures were simulated with the finite difference time domain method (FDTD Solutions package from Lumerical Inc). The modeling of near field-driven photocurrents in graphene was implemented by our Python codes which solve the Navier–Stokes equations.

**Device fabrication.** We started the fabrication from a thermally oxidized (~ 285 nm) highly resistive Si wafer (10 kΩ·cm, Nova electronic materials). Few-layer graphene flakes were mechanically exfoliated from natural graphite (NGS Naturgraphit) onto the wafer and identified with an optical microscope. Metallic alignment markers were fabricated with electron-beam lithography (EBL, JBX-6300FS, Jeol), Ebeam evaporation of 3/20nm Ti/Au (AJA international inc.), and liftoff in hot acetone at 65°C for 1 hour. The recipe of our EBL process is spin-coating of PMMA 495k A5 at 4000 rpm; baking at 180°C for 2 mins; exposure dose of 1300 μC/cm$^2$; development in MIBK: IPA=1:3 for 30 s and rinse in IPA for 30 s; drying with a nitrogen gun. After that, we patterned the graphene sheet to desired shapes such as ring, half ring, L-shaped cascaded devices, and ribbons, using a second EBL and oxygen plasma etching (20 sccm O$_2$, 20W, 20 s, VITA, Femto Science Inc). The samples were then annealed at 300 °C for 6 hours in Ar/H$_2$ atmosphere to remove the resist residues. Afterward, plasmonic nanostructures and contact electrodes were patterned on graphene flakes by a third EBL with an alignment accuracy of about 100 nm, followed by thermal deposition (5/60 nm Pd/Au, Kurt J. Lesker company) and liftoff in hot acetone at 65°C for 1 hour. A fourth EBL and thermal deposition (5/80 nm Cr/Au) were used to fabricate the large electrodes for probing and wire bonding. Finally, we deposit a thin 12 nm thick Al$_2$O$_3$ to encapsulate the device, which consists of two steps: firstly, the deposition of 2 nm metal Al with a high-vacuum Ebeam evaporator which will be then quickly oxidized to form uniform Al$_2$O$_3$ film on graphene; secondly, the deposition of 10 nm Al$_2$O$_3$ with ALD. The Al$_2$O$_3$ layer helps reduce the doping level of the graphene device and hence increases the responsivity. The existence of the thin Al$_2$O$_3$ layer does not affect our probing (5 μm radius tip) and wire bonding.

**Characterization**. We used a quantum cascade laser at 4 μm wavelength (MIRCat-1200, Daylight Solutions, polarization ratio > 100:1) as our light source. A low-order quarter-wave plate (WPLQ05M-4000, Thorlabs) designed at 4 μm was used to control the polarization states of light. The light was delivered onto the device by plane mirrors and an off-axis parabolic mirror (focal length=101.6 mm, MPD149-P01, Thorlabs). Normal incidence was ascertained before experiments. The open-circuit voltages were measured as the photoresponse with a lock-in amplifier (Stanford research systems, SR830), and the light signal was modulated by an optical chopper (Stanford research systems, SR540, up to 4 kHz modulation frequency) at about 280 Hz to get rid of the low-frequency noises and increase the signal-to-noise ratio. We used a filter slope of 24 dB/oct, and the resulting equivalent noise bandwidth (ENBW) of the lock-in amplifier is then $5/(64t)$, where $t$ is the time constant. We used a time constant of 30 ms for noise measurement, corresponding to an ENBW of 2.6 Hz. The d.c. *I-V* curves and gate voltage dependence were measured with a semiconductor characterization system (Keithley 4200-SCS) without preamplifiers. The characterization of on-off cycles and response time were conducted by directly connecting the device to an oscilloscope (InfiniiVision DSOX3034T, Keysight, 350 MHz, 5 GSa/s). The average mode was used to increase the signal-to-noise ratio. The 1000 ns pulse was generated by our QCL laser in pulse mode. The power level of incident light was tuned with a set of neutral density filters and calibrated with a power meter (843-R, Newport).

# Data availability

All data needed to evaluate the conclusions in this paper are present in the paper or the Supplementary Materials. Additional data related to this paper may be requested from the corresponding authors C.L. or C.-W.Q. upon request.